\documentclass{elsart}
\usepackage{amssymb}
\usepackage[dvips]{graphicx}
\usepackage{amssymb}
\usepackage{amsmath}

\input{tcilatex}
\journal {\textbf{European Journal of Mechanics},\textbf{ B/Fluids}, vol. 9, N$^o$ 5, pp. 469-491 (1990).\quad\qquad\quad\qquad\quad\qquad\quad\quad\qquad\quad\qquad\qquad}
\input{tcilatex}
\begin{document}

\begin{frontmatter}

\title{ Variational Theory of Mixtures \\ in Continuum Mechanics}

\author{Henri Gouin}
\ead{henri.gouin@univ-cezanne.fr}

\address {
C.N.R.S. \ U.M.R. 6181 \&  Universit\'e d'Aix-Marseille     \\ Case
322, Av. Escadrille
 Normandie-Niemen, 13397 Marseille Cedex 20 France}

\begin{abstract}
In continuum mechanics, the equations of motion for mixtures were derived
through the use of a variational principle by Bedford and Drumheller in
[1978]. Only immiscible mixtures were investigated. We have chosen a
different approach.

In this paper, we first write the equations of motion for each
constituent of an inviscid miscible mixture of fluids without
chemical reactions or diffusion. The theory is based on Hamilton's
extended principle and regards the mixture as a collection of
distinct continua.\\
 The internal energy is assumed to be a function
of densities, entropies and successive spatial gradients of each
constituent. Our work leads to the equations of motion in an
universal thermodynamic form in which interaction terms subject to
constitutive laws, difficult to interpret physically, do not occur.
\\
For an internal energy function of densities, entropies and spatial
gradients, an equation describing the barycentric motion of the constituents
is obtained.
\\
The result is extended for dissipative mixtures and an equation of energy is
obtained. A form of Clausius-Duhem's inequality which represents the second
law of thermodynamics is deduced. In the particular case of compressible
mixtures, the equations reproduce the classical results.
\\
Far from critical conditions, the interfaces between different phases in a
mixture of fluids are layers with strong gradients of density and entropy.
The surface tension of such interfaces is interpreted.
\end{abstract}


\end{frontmatter}

\section{Introduction}

The knowledge of equations describing processes occuring in fluid mixtures
in one or several phases is scientifically and industrially very important
([Barrere \& Prud'homme, 1973]; [Ishii, 1975]). Many papers concerning
theoretical and experimental works for mixtures have been produced in
continuum mechanics and physical chemistry. References in [Bedford \&
Drumheller, 1983] cover most of the works in continuum theory. The book by
Rowlinson \& Swinton [1982] gives the statistical methods.
\\
Truesdell [1957,1965] first derived equations of balance and motion
through the use of a continuum theory of mixtures. The mixture is
then considered as a distribution of different continuous media in
the same physical space, at time \textit{t}. Each constituent of the
mixture is identified with its reference space. The thermal and
mechanical equations of balance are introduced for each constituent.
They keep the usual form for a fluid description but they must
include numerous terms of interaction between the different
constituents. The study of the average motion, the motion of the
barycentre of constituents with coefficients equal to their
respective densities, are deduced from the sum of momenta, energy
and entropy of each constituent. The terms including interaction
between different constituents of the mixtures are difficult to
interpret physically. They require constitutive postulates that are
difficult to interpret experimentally.
\\
Several assumptions are open to doubt. How must we choose an
entropy? (Mixture entropy or entropy for each constituent?) What is
the form of the second law of thermodynamics? Is it symbolized with
one or several Clausius-Duhem's inequalities? Must we consider a
temperature of the mixture or a temperature for each constituent?
\\
M\"{u}ller [1967]; Williams [1973]; Atkin \& Craine [1976]; Sampio \&
Williams [1977]; Bowen [1979]; Nunziato \& Walsh [1980]; Drew [1983] and
many others argue about the various points of view.
\\
In several papers Bedford and Drumheller [1978]; [B \& D, 1979]; [B
\& D, 1980] presented a variational theory for immiscible mixtures
(a continuum theory for mixtures whose constituents remain
physically separated, such as a mixture of immiscible liquids or a
fluid containing a distribution of particles, droplets, or bubbles).
That the constituents remain physically separated has several
implications for the theory. For instance, the authors use a
variational principle in which a variation is added to the motion of
only one constituent. However it is necessary to add to the
variations of kinetic and potential energies a virtual work due to
the forces not represented by a potential. Such forces are difficult
to take into account in a variational principle deduced from the one
by Hamilton. It is not possible to use such a method in a more
general model.
\\
In this paper a new mathematical systematic method that leads to the
equations of motion and energy for miscible fluid mixtures without chemical
reaction and diffusion is propounded. The presented work is essentially
different from the one by Bedford and Drumheller.
\\
The mixture is represented by several distinct continuous media that
occupy the same physical space at time \textit{t}. The novelty
consists in the fact that the used variational principle is applied
to a lagrangian representation associated with a reference space for
each component. The proposed method uses the knowledge of the
internal energy of the mixture at every point of the physical space.
This is not the case in [B \& D].
\\
For classical fluids, the method corresponds to Hamilton's principle in
which one makes a variation of the reference position of the particle. It
was proposed by Gouin [1987].
\\
The inferred equations are written in a different form that facilitates the
study of first integrals for conservative motions ( [Casal, 1966]; [Gouin,
1981]; [Casal \& Gouin, 1985\textit{b}, 1988\textit{a}, 1989]).
\\
For mixtures, the principle appears far more convenient than the method of
only taking the variation of the average motion into consideration: it
separately tests each component and leads to a thermodynamic form of the
equation of motion for each of them.
\\
The internal energy is assumed to be a function of different densities or
concentrations of the mixture. Because of the form of the principle we use,
one must consider an entropy for each component in the mixture. The total
entropy of the system is the sum of the partial entropies.
\\
To consider areas where strong gradients of density occur - for example
shocks or capillary layers - the internal energy is chosen as a function of
the successive derivatives of densities and entropies.
\\
Two cases of conservative motions for mixtures are considered: isentropic
motion and isothermal motion. These are only limiting mathematical cases for
testing the method. The physical motion is intermediate, between these
extremes.
\\
An important medium is the one of \textit{thermocapillary mixtures}. It
corresponds to an energy taking into account the gradients of densities and
entropies. Then, one obtains an equation of the barycentric motion of
constituents. For a compressible mixture, the equations of motion involve a
hydrostatic stress.
\\
In the same way as for thermocapillary fluids, an additional term
that has the physical dimension of a heat flux may be added to the
energy equation ([Eglit, 1965]; [Berdichevskii, 1966]; [Casal \&
Gouin, 1985\textit{a}]; [Dunn \& Serrin, 1985]; [Gatignol \&
Seppecher, 1986]; [Seppecher, 1987]). A generalization of the
Clausius-Duhem inequality is obtained for non-conservative mixtures.
The inequality is deduced from the heat conduction inequality by
introducing an irreversible stress tensor associated with a general
dissipation function. The theory fits with the second law of
thermodynamics.
\\
From these results one generalizes equilibrium conditions obtained through
interfaces in compressible fluid mixtures [Rocard, 1967]. An interpretation
of capillarity stresses is deduced (rule of Antonov) [Emschwiller, 1964],
and generalization of results by Vignes-Adler \& Brenner [1985] may be
investigated.

\section{The motion of a continuous medium}

\subsection{Motion of a fluid with only one constituent}

Recall that the motion of a continuous medium can be represented by a
surjective differentiable mapping:%
\begin{equation*}
\emph{\textbf{z}}\rightarrow \
\emph{\textbf{X}}=\mathbf{M}(\emph{\textbf{z}}),
\end{equation*}%
where $\emph{\textbf{z}}=(t,\emph{\textbf{x}})$ belongs to
$\mathcal{W}$ , an open set in the time-space occupied by the fluid
between time $t_{1}$ and time $t_{2}$. The
position of a particle in a reference space $\mathcal{D}_{o}$ is denoted by $%
\emph{\textbf{X}}$; its position at time $t$ in $\mathcal{D}_{t}$ is denoted by $%
\emph{\textbf{x}}$ [Serrin, 1959]. Hamilton's principle -
variational form of the principle of virtual powers - allows us to
investigate the equation of motion.

The variations of motion of particles are deduced from:%
\begin{equation*}
\emph{\textbf{X}}=\mathbf{\Psi }(\emph{\textbf{x}},t;\beta )
\end{equation*}%
for which $\beta $ is a parameter defined in a neighborhood of zero; it is
associated with a family of virtual motions of the fluid. The real motion
corresponds to $\beta =0$.
\\
Virtual material displacements associated with any variation of real
motion can be written [Gouin, 1987]:
\begin{equation}
\delta \emph{\textbf{X}}= \left.\frac {\partial \mathbf{\Psi }}{\partial \beta }%
\right\vert_{\beta =0}.  \tag{1}
\end{equation}%
Such a variation is dual with Serrin's [S, 1959]. This has been studied in
the case of compressible perfect fluids [Gouin, 1978] and corresponds to the
natural variation of the motion in a lagrangian representation.

The lagrangian of the fluid is:%
\begin{equation*}
L=\rho (\frac{1}{2}\mathbf{V}^{\ast }\mathbf{V}-\alpha -\Omega ),
\end{equation*}
where $\mathbf{V}$ denotes the velocity vector of the particles , $\Omega $
the extraneous force potential defined in $\mathcal{D}_{t}$ , $\alpha $ the
specific internal energy, $\rho $ the density and $^{\ast }$ the
transposition in $\mathcal{D}_{t}$.

Between time $t_{1}$ and time $t_{2}$, the hamiltonian action is written:
\begin{equation}
a=\int_{t_{1}}^{t_{2}}\int_{\mathcal{D}_{t}}L\,dvdt.  \tag{2}
\end{equation}
The use of virtual displacements (1) provides the equations of motion of
compressible perfect fluids in a thermodynamic form. For fluids of grade $n$%
, the calculus is swift. It generalizes the equation (29,8) of
fluids   in [Serrin, 1959] that may be written in a form that is
independent of their complexity [Gouin, 1987]:
\begin{equation*}
\mathbf{\Gamma }=\theta \func{grad}s-\func{grad}(h+\Omega ),
\end{equation*}%
where $\mathbf{\Gamma }$ denotes the acceleration, $s$ the entropy, $\theta $
a temperature and $h$ a specific enthalpy.

\subsection{Motion of a fluid mixture}

For the sake of clarity, we study a mixture of two fluids. The
method can be immediately extended to any number of constituents. No
assumption has to be made about their composition or their
miscibility.
\\
The motion of a two-fluid continuum can be represented using two surjective
differentiable mappings:%
\begin{equation}
\begin{array}{lllll}
\emph{\textbf{z}}\rightarrow
\emph{\textbf{X}}_{1}=\mathbf{M}_{1}(\emph{\textbf{z}}) &  &
\text{and} & &
\emph{\textbf{z}}\rightarrow \ \emph{\textbf{X}}_{2}=\mathbf{M}_{2}(\emph{\textbf{z}}),%
\end{array}
\tag{3}
\end{equation}%
(Subscripts 1 and 2 are associated with each constituent).

\begin{figure}[h]
\begin{center}
\includegraphics[width=8cm]{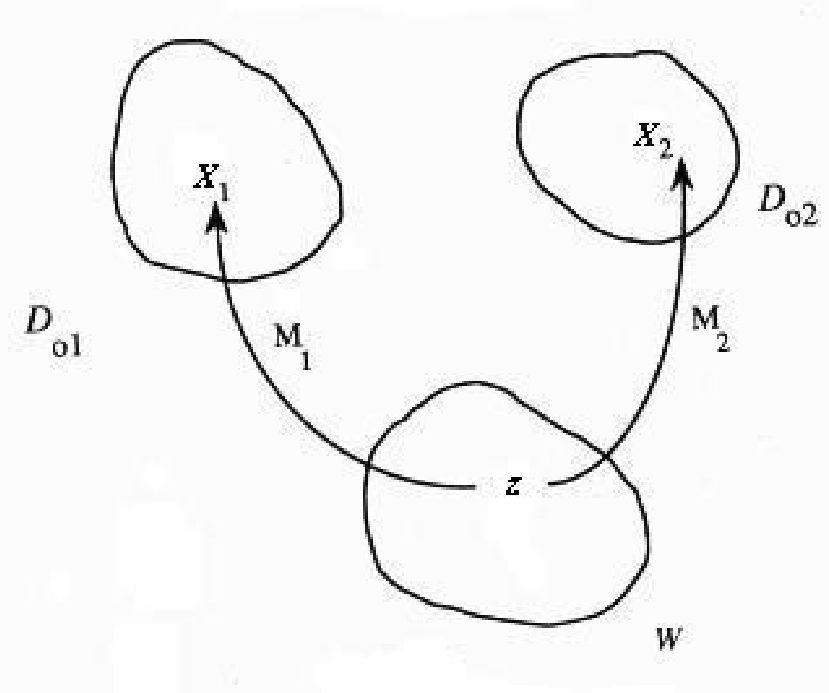}
\end{center}
\caption{}\label{fig1}
\end{figure}

$\emph{\textbf{X}}_{1}$ and $\emph{\textbf{X}}_{2}$ denote the
positions of each constituent in reference spaces $\mathcal{D}_{o1}$
and $\mathcal{D}_{o2}$. The variations of motion of particles are
deduced from:
\begin{equation}
\begin{array}{lllll}
\emph{\textbf{X}}_{1}=\mathbf{\Psi }_{1}(\emph{\textbf{x}},t;\beta
_{1}) & & \text{and} &
& \emph{\textbf{X}}_{2}=\mathbf{\Psi }_{2}(\emph{\textbf{x}},t;\beta _{2}).%
\end{array}
\tag{4}
\end{equation}
This is a generalization of the one in paragraph 2.1 where $\beta _{1}$ and $%
\beta _{2}$ are defined in a neighborhood of zero; they are associated with
a two-parameter family of virtual motions of the mixture. The real motion
corresponds to $\beta _{1}$ $=0$ and $\beta _{2}$ $=0$; the associated
virtual displacements can be written:
\begin{equation}
\begin{array}{lllll}
\displaystyle \delta _{1}\emph{\textbf{X}}_{1}=\left . \frac
{\partial \mathbf{\Psi }_{1}}{\partial \beta _{1}}\right\vert_{\beta
_{1}=0} &  & \text{and} &  &
\delta _{2}\emph{\textbf{X}}_{2}=\left . \displaystyle\frac {\partial \mathbf{\Psi }%
_{2}}{\partial \beta _{2}}\right\vert_{\beta _{2}=0}.%
\end{array}
\tag{5}
\end{equation}
They generalize relation (1) given for a fluid with one constituent. The
principle tests separately each constituent of the mixture and so is very
efficient.

\textit{\quad Remark}. - It follows from the classical
representation of a
two-constituent mixture, that formula (4) can be written%
\begin{equation}
\begin{array}{lllll}
\emph{\textbf{x}}=\mathbf{\varphi
}_{1}(\emph{\textbf{X}}_{1},t;\beta _{1}) &  & \text{and}
&  & \emph{\textbf{x}}=\mathbf{\varphi }_{2}(\emph{\textbf{X}}_{2},t;\beta _{2}).%
\end{array}
\tag{6}
\end{equation}
A variation of the motion like Serrin's (p. 145, [S, 1959]) cannot
be extended to any two-parameter family of virtual displacements: as
a matter of fact, any field of virtual displacements associated with
$\mathbf{\varphi }_{1}$ must be such that $\delta \mathbf{x}=$
$\displaystyle\frac {\partial \mathbf{\varphi }_{1}}{\partial \beta
_{1}}(\emph{\textbf{X}}_{1},t;0)$ and there
is no reason for being identifiable with a vector field written in the form $%
\delta \mathbf{x}=\displaystyle \frac {\partial \mathbf{\varphi }_{2}}{%
\partial \beta _{2}}(\emph{\textbf{X}}_{2},t;0).$
\\
In the case of compressible fluids, writing (1) for the virtual displacement
associated with a one-parameter family of virtual motion of the fluid is
deduced from the variation defined by Serrin by the way of vector spaces
isomorphism [Gouin, 1976]. \textit{For mixtures, by using }(5)\textit{, a
virtual displacement of miscible fluid mixtures can be considered}.

The lagrangian of the mixture is:%
\begin{equation*}
L=\frac{1}{2}\rho _{1}\mathbf{V}_{1}^{\ast }\mathbf{V}_{1}+\frac{1}{2}\rho
_{2}\mathbf{V}_{2}^{\ast }\mathbf{V}_{2}-\varepsilon -\rho _{1}\Omega
_{1}-\rho _{2}\Omega _{2}.
\end{equation*}%
Here, $\mathbf{V}_{1}$ et $\mathbf{V}_{2}$ denote the velocity
vectors of each constituent, $\rho _{1}$ and $\rho _{2}$ are the
densities, $\Omega _{1} $ and $\Omega _{2}$ are the extraneous force
potentials depending only on
$\emph{\textbf{z}}=(t,\emph{\textbf{x}})$ and $\varepsilon $ is the
internal energy per unit volume.
\\
The expansion of the lagrangian is general. In fact dissipative
phenomena imply that $\mathbf{V}_{1}$ is almost equal to
$\mathbf{V}_{2}$. Because of the interaction of the constituents,
$\varepsilon $ does not always divide  into energies related to each
constituent of the mixture, as it does in the case of \textit{simple
mixtures of fluids} [M\"{u}ller, 1968].
\\
More generally, the lagrangian may be written%
\begin{equation*}
L=\frac{1}{2}\rho _{1}\mathbf{V}_{1}^{\ast }\mathbf{V}_{1}+\frac{1}{2}\rho
_{2}\mathbf{V}_{2}^{\ast }\mathbf{V}_{2}-\eta
\end{equation*}
where $\eta $ denotes the volumetric potential energy of the
mixture. In our particular case, $\eta $ is the sum of internal
energy and extraneous force potentials. In fact, the following
calculus would be similar (but even simpler). We use the first form
of the lagrangian in order to obtain classical expressions used for
fluids.

Hamilton's action between time $t_{1}$ and time $t_{2}$ can be
written as in (2). The mixture is assumed  not to be chemically
reacting. The conservation of
 masses requires that:%
\begin{equation}
\rho _{i}\det F_{i}=\rho _{oi}(\emph{\textbf{X}}_{i})  \tag{$ 7^{1}
$}
\end{equation}%
for the densities of each constituent,\ with subscript $i$ belonging to $%
\{1,2\}$. At fixed $t$, the jacobian mapping associated with each
$\mathbf{M}_{i}$ is
denoted by $F_{i}$; $\rho _{oi}$ is the reference specific mass in $\mathcal{%
D}_{oi}$.
\\
In differentiable cases, equation $(7^{1})$ is equivalent to:%
\begin{equation}
\frac{\partial \rho _{i}}{\partial t}+\func{div}\rho _{i}\mathbf{V}_{i}=0.
\tag{$7^{2}$}
\end{equation}%
The volumetric internal energy $\varepsilon $ is given by the
thermodynamical behavior of the mixture and will appear in the equations of
motion. Each constituent has a specific mass; similarly, two specific
entropies $s_{1}$ and $s_{2}$ are supposed to be associated with
constituents 1 and 2.
\\
For a specific internal energy depending on derivatives of gradients
of densities and entropies up to the $n$-th order:
\begin{eqnarray*}
\varepsilon &=&\varepsilon (s_{1},\func{grad}s_{1},...,(\func{grad}%
)^{n-1}s_{1};s_{2},\func{grad}s_{2},...,(\func{grad})^{n-1}s_{2}; \\
&&\rho _{1},\func{grad}\rho _{1},...,(\func{grad})^{n-1}\rho _{1};\rho _{2},%
\func{grad}\rho _{2},...,(\func{grad})^{n-1}\rho _{2}), \qquad\qquad\quad (8)
\end{eqnarray*}%
(we say, the mixture is of grade $n$), the calculus is similar to
than in ([Gouin, 1987]; [Casal \& Gouin, 1988\textit{a}]). The
orders of gradients of $\rho _{1}$, $\rho _{2}$, $s_{1}$, $s_{2}$
can be chosen differently, but the exposition would be less clear.
Let us note that the volumetric potential energy can be written:
\begin{eqnarray*}
\eta &=&\eta (s_{1},\func{grad}s_{1},...,(\func{grad})^{n-1}s_{1};s_{2},%
\func{grad}s_{2},...,(\func{grad})^{n-1}s_{2}; \\
&&\rho _{1},\func{grad}\rho _{1},...,(\func{grad})^{n-1}\rho _{1};\rho _{2},%
\func{grad}\rho _{2},...,(\func{grad})^{n-1}\rho
_{2},\emph{\textbf{z}})
\end{eqnarray*}%
with
$
\emph{\textbf{z}}=(t,\emph{\textbf{x}}).
$
\subsection{Examples of motions of a fluid mixture}
Let us consider two examples of Hamilton's principle. Obviously, they are
two limit cases deduced from a variational principle and incompatible with
any dissipative process.

\textit{{\quad First case}:} Isentropic motions.

We assume that the specific entropy of each particle is constant:%
\begin{equation*}
\begin{array}{lllll}
\displaystyle \frac {ds_{i}}{dt}=0, &  & \text{or} &  & \displaystyle\frac {%
\partial s_{i}}{\partial t}+\frac {\partial s_{i}}{\partial \emph{\textbf{x}}}%
\mathbf{V}_{i}=0.%
\end{array}%
\end{equation*}%
(Lie's  derivative with respect to the velocity field of constituent
$i$ is equal to zero. The average motion of the mixture does not
have Lie's derivative, so henceforth, we can expect each constituent
of the mixture to possess entropy).
\\
The equation
\begin{equation}
s_{i}=s_{oi}(\emph{\textbf{X}}_i)  \tag{9}
\end{equation}%
defines an isentropic motion.
\\
Motion equations are deduced from Hamilton's principle. The two variations
of hamiltonian action are:%
\begin{equation*}
\delta _{i}a=a^{\prime }(\beta _{i})\left\vert _{\beta _{i}=0}\right. .
\end{equation*}%
For $i\in \{1,2\}$, we deduce:
\begin{eqnarray*}
\delta _{i}a &=&\int_{t_{1}}^{t_{2}}\int_{\mathcal{D}_{t}}\left\{ \left(\frac{1}{2%
}\mathbf{V}_{i}^{\ast }\mathbf{V}_{i}-\varepsilon _{,\rho _{i}}-\Omega
_{i}\right)\delta _{i}\rho _{i}+\rho _{i}\mathbf{V}_{i}^{\ast }\delta _{i}\mathbf{V%
}_{i}-\varepsilon _{,s_{i}}\delta _{i}s_{i}\right. - \\
&&\qquad\qquad \left[\varepsilon _{,s_{i,_{\gamma }}}\delta
_{i}s_{i_{,\gamma }}+...+\varepsilon _{,s_{i,_{\gamma
_{1},...,\gamma _{n-1}}}}\delta
_{i}s_{i_{,\gamma _{1},...,\gamma _{n-1}}}\right]- \\
&&\qquad\qquad \left. \left[ \varepsilon _{,\rho _{i,_{\gamma
}}}\delta _{i}\rho _{i_{,\gamma }}+...+\varepsilon _{,\rho
_{i,_{\gamma _{1},...,\gamma _{n-1}}}}\delta _{i}\rho _{i_{,\gamma
_{1},...,\gamma _{n-1}}}\right]\right\} dvdt.
\end{eqnarray*}
Subscript $\gamma $ corresponds to spatial derivatives associated with
gradient terms; as usually, summation is made on repeated subscripts $\gamma
$.

The quantity%
\begin{equation}
h_{i}=\varepsilon _{,\rho _{i}}+\sum_{p=1}^{n-1}(-1)^{p}(\varepsilon _{,\rho
_{i,_{\gamma _{_{1}},...,\gamma _{_{p}}}}})_{,_{\gamma _{_{1}},...,\gamma
_{p}}}  \tag{$10^{1}$}
\end{equation}%
\textit{defines the }specific enthalpy\textit{\ of the constituent }$i$%
\textit{\ of the mixture}, and
\begin{equation}
\theta _{i}=\frac{1}{\rho _{i}}\left[\varepsilon
_{,s_{i}}+\sum_{p=1}^{n-1}(-1)^{p}(\varepsilon _{,s_{i,_{\gamma
_{_{1}},...,\gamma _{_{p}}}}})_{,_{\gamma _{_{1}},...,\gamma
_{p}}}\right] \tag{$10^{2}$}
\end{equation}%
\textit{defines the }temperature\textit{\ of the constituent }$i$\textit{\
of the mixture.}

The motion of constituent $i$ of the mixture is expressed by the equation
\begin{equation}
\mathbf{\Gamma }_{i}=\theta _{i}\func{grad}s_{i}-\func{grad}(h_{i}+\Omega
_{i}).  \tag{11}
\end{equation}%
\textit{Appendix 1 gives details of the calculation.}

The fact that an entropy is introduced for each component implies the
existence of a temperature for each fluid constituent. This is in accordance
with [Green and Naghdi, 1965]. In this first limit case, the variational
principle does not lead to the equality of temperatures. The behavior of the
two components corresponds to their thermodynamical independence (no
exchange of heat between particles). The two components are only bounded by
effects of compressibility. Such a model can be used in practice only if one
includes the fact that some heat conduction exists and adjusts the values of
the two temperatures. So, one verifies that equations obtained with the
variational principle must be completed.

\textit{{\quad Second case}:} Only the whole entropy of the mixture
is conserved.

We have:%
\begin{equation}
\int_{\mathcal{D}_{t}}\left(\rho _{1}s_{1}+\rho
_{2}s_{2}\right)dv=K, \tag{12$^{1}$}
\end{equation}
where $K$ is constant (independent of the time).

Let us use the method applied in [Casal \& Gouin, 1988\textit{b}]
for compressible fluids. The variation of the entropy of each
component is assumed to be the sum of a variation associated with
the virtual motion and
another one, $\delta _{oi}s_{i}$ related to the particle:%
\begin{equation*}
\delta _{i}s_{i}=\frac{\partial s_{oi}}{\partial \emph{\textbf{X}}_{i}}\,\delta _{i}%
\emph{\textbf{X}}_{i}+\delta _{oi}s_{i},
\end{equation*}%
where $\delta _{oi}s_{i}$ and $\delta _{i}\emph{\textbf{X}}_{i}$ are
independent
and $\delta _{oi}$ must verify, with respect to component $i$ of the mixture:%
\begin{equation}
\delta _{oi}\int_{t_{1}}^{t_{2}}\int_{\mathcal{D}_{t}}(\rho _{1}s_{1}+\rho
_{2}s_{2})dv=0.  \tag{12$^{2}$}
\end{equation}%
Classical methods of variational calculus provide the variation of
the hamiltonian action: the variations $\delta
_{i}\emph{\textbf{X}}_{i}$ yield a motion equation of the form of
(11).
\\
Relation (12$^{2}$) implies: there exists a constant Lagrange multiplier $%
\theta _{o}$, \textit{constant} \textit{in space and in time}, such that:
\begin{eqnarray*}
\delta _{oi}a
&=&\int_{t_{1}}^{t_{2}}\int_{\mathcal{D}_{t}}\left(-\varepsilon
_{,s_{i}}\delta _{oi}s_{i}-\varepsilon _{,s_{i,_{\gamma }}}\delta
_{oi}s_{i_{,\gamma }}-... \right.\\
&&\left.\qquad\qquad\qquad\qquad-\varepsilon _{,s_{i,_{\gamma
_{1},...,\gamma _{n-1}}}}\delta _{oi}s_{i_{,\gamma _{1},...,\gamma
_{n-1}}}+\rho _{i}\theta _{o}\delta _{oi}s_{i} \right)dvdt.
\end{eqnarray*}%
Let us point out that adiabatic condition (12$^{1}$) implies that even if
the mixture is moving, $\theta _{o}$ remains constant in space and time
since $\theta _{o}$ is a Lagrange multiplier associated with an integral
constraint.

Considering variations $\delta _{oi}s_{i}$ vanishing on the boundary of $%
\mathcal{D}_{t}$ ; an integration by parts yields:
\begin{equation*}
\delta _{oi}a=\int_{t_{1}}^{t_{2}}\int_{\mathcal{D}_{t}}\rho
_{i}(\theta _{o}-\theta _{i})\,\delta _{oi}s_{i}\,dvdt.
\end{equation*}%
The principle: "for any $\delta _{oi}s_{i}$ vanishing on the boundary of $%
\mathcal{D}_{t}$ , $\delta _{oi}a=0$ " leads to the equation for the
temperature:
\begin{equation}
\theta _{i}=\theta _{o},  \tag{13}
\end{equation}%
\textit{i.e., }all the constituents have the same temperature $\theta _{o}$,
constant in all the flow: the flow is \textit{isothermal}.

This variational principle can be compared with the static one expressing
that the equilibrium of a compressible fluid kept in a fixed adiabatic
reservoir without exchange of mass and entropy with outside is obtained by
writing the uniformity of temperature and pressure. It can be interpreted as
a limit case of a mixture with an infinite heat conductivity.
\\
Equation (11) is replaced by
\begin{equation}
\mathbf{\Gamma }_{i}+\func{grad}(g_{i}+\Omega _{i})=0,  \tag{14}
\end{equation}
where $g_{i}=h_{i}-\theta _{i}s_{i}$ is the free specific enthalpy (or
chemical potential) of constituent $i$.
\\
When the fluid is compressible, Eqs (11) or (14) yield first integrals and
theorems of circulation [Serrin, 1959]; [Casal, 1966]; [Gouin, 1981]; [Casal
\& Gouin, 1985\textit{b}, 1989]. These properties can easily be extended to
each constituent of an inviscid mixture.
\\
Each constituent of the mixture in an isothermal equilibrium state satisfies
\begin{equation}
G_{i}=C_{i}  \tag{15}
\end{equation}%
where $G_{i}=g_{i}+\Omega _{i}$ and $C_{i}$ is a constant.

An interface in a two-phase mixture is generally modelled by a surface
without thickness. Far from critical conditions, this layer is of molecular
size; density and entropy gradients are very large. A continuous model deals
with such areas by using energy in the form (8) extending forms given in
([Rocard, 1967]; [Cahn \& Hilliard, 1959]) for compressible fluids.

Equation (15) is satisfied for each constituent of a mixture in an
isothermal equilibrium state. The fact that $C_{i}$ values are the
same in each bulk is also a consequence of Eq.(15). It follows that
with extraneous forces being neglected in a two-phase mixture, if
the two different phases are in isothermal equilibrium, then
specific free enthalpies (or chemical potentials) of each bulk are
equal. This result generalizes the classical law to mixtures of
grade $n$: the different constituents keep their own chemical
potential through a change of phases [Rocard, 1967].

\section{Thermocapillary mixtures}

The previous equations yield the equations of motion for each of the
two fluids constituting the mixture. Usually, the chemical
potentials $h_{i}$ and $g_{i}$ are rarely considered except for
equilibrium; Eqs. (11) and (14) make use of them in the study of the
motion.
\\
In order to compare results given by the variational technics and
the usual approach of mixture motions, we can try to study the mean
motion and diffusive motions for two components. But how do we
define these motions? In this paper, we consider the mean motion
using an approach which is independent of the number of
constituents. \\
For the mean motion, $\rho =\sum_{i}\rho _{i}$ and $\rho\; \mathbf{\Gamma }%
=\sum_{i}\rho _{i}\mathbf{\Gamma }_{i}$  (using $\mathbf{\Gamma }=\sum_{i}%
\mathbf{\Gamma }_{i}$ has no physical meaning).
\\
On the contrary, $\rho\; \mathbf{\Gamma }$ has the physical
dimension of a volumetric force and is an additive quantity in the
same way as body forces or the stress tensor divergences. So, in
this first article, we have not attempted to investigate the
difficult question of the relative motion associated with "
$\mathbf{\Gamma }_{1}-\mathbf{\Gamma }_{2}$ " and hence to see how
diffusion behaves in miscible mixtures.

\subsection{Conservative motions for mixtures of thermocapillary fluids}

In the case $n=2$, we write the equations of motion by using the stress
tensor. The mixture is said to be a \textit{thermocapillary mixture} if the
internal energy $\varepsilon $ is written as:
\begin{equation*}
\varepsilon =\varepsilon
(s_{1},\func{grad}s_{1},s_{2},\func{grad}s_{2},\rho
_{1},\func{grad}\rho _{1},\rho _{2},\func{grad}\rho _{2}).  \tag{16}
\end{equation*}
The expressions
\begin{equation*}
\begin{array}{lllll}
\rho =\displaystyle \sum_{i=1}^{2}\rho _{i} &  & \text{and} &  &\displaystyle \rho\; \mathbf{\Gamma }%
=\sum_{i=1}^{2}\rho _{i}\mathbf{\Gamma }_{i}%
\end{array}%
\end{equation*}%
introduce the total density and acceleration of the mixture. With
denoting by $\alpha $ the specific energy of the mixture, we have
\begin{equation*}
\rho \alpha =\varepsilon ,
\end{equation*}%
and,
\begin{equation*}
d\alpha =\sum_{i=1}^{2}\frac{\mathcal{P}_{i}}{\rho \rho _{i}}\,d\rho _{i}+%
\frac{\rho _{i}}{\rho }\Theta _{i}\;ds_{i}+\frac{1}{\rho }\Phi
_{i\gamma }d\rho _{i,\gamma }+\frac{1}{\rho }\Psi _{i\gamma
}ds_{i,\gamma },
\end{equation*}%
with
\begin{equation*}
\begin{array}{lllllll}
\mathcal{P}_{i}=\rho \rho _{i}\alpha _{,\rho _{i}}, &  & \Theta _{i}=%
\displaystyle \frac{1}{\rho _{i}}\,\varepsilon _{,s_{i}}, &  & \Phi
_{i\gamma }=\varepsilon _{,\rho _{i,\gamma }}, &  & \Psi _{i\gamma
}=\varepsilon
_{,s_{i,\gamma }}.%
\end{array}%
\end{equation*}%
Four new vectors $\mathbf{\Phi }_{i}$ and $\mathbf{\Psi }_{i}$ $(i\in
\{1,2\})$ are introduced by the theory; it is an extension of the calculus
written in [Casal \& Gouin, 1988\textit{a}].
\\
For a compressible fluid (with only one constituent), $\mathcal{P}$ and $%
\Theta $ are respectively the pressure and the Kelvin temperature. The fluid
is isotropic, so the internal energy is of the form
\begin{equation*}
\varepsilon =\varepsilon (\rho _{i},s_{i},\beta _{ij},\chi _{ij},\gamma
_{ij})
\end{equation*}%
with $ \beta _{ij}=\func{grad}\rho _{i}.\func{grad}\rho _{j}$, $\chi _{ij}=%
\func{grad}\rho _{i}.\func{grad}s_{j}$, $\gamma _{ij}=\func{grad}s_{i}.\func{%
grad}s_{j}$ $(i\in \{1,2\})$.
\\
It follows\ \ that:%
\begin{equation*}
\mathbf{\Phi }_{i}=\sum_{j=1}^{2}C_{ij}\func{grad}\rho _{j}+E_{ij}\func{grad}%
s_{j},
\end{equation*}%
\begin{equation*}
\ \  \mathbf{\Psi }_{i}=\sum_{j=1}^{2}D_{ij}\func{grad}s_{j}+E_{ij}\func{grad}%
\rho _{j},
\end{equation*}%
with
\begin{equation*}
\begin{array}{lllll}
C_{ij}=(1+\delta _{ij})\varepsilon _{,\beta _{ij}}, &  & D_{ij}=(1+\delta
_{ij})\varepsilon _{,\gamma _{ij}}, &  & E_{ij}=\varepsilon _{,\chi _{ij}},%
\end{array}%
\end{equation*}%
where $\delta _{ij}$ is the Kronecker symbol. The simplest model is
for $C_{ij}$, $D_{ij}$ and $E_{ij}$ constant.

Assuming that $\Omega _{1}=\Omega _{2}=\Omega $  (this is the case
for gravity forces), we obtain a general formulation of the equation
of motion
for thermocapillary mixtures:%
\begin{equation*}
\rho\, \mathbf{\Gamma }=\func{div}\sigma -\rho \func{grad}\Omega
 \tag{17}
\end{equation*}%
or
\begin{equation*}
\rho\, \Gamma _{\gamma }=\sigma _{\nu \gamma _{,\nu }}-\rho\, \Omega
_{,\gamma }.
\end{equation*}%
The sum of $\sigma _{1}$ and $\sigma _{2}$ obtained by successive
integrations by parts is conventionally called the generalization of the
stress tensor. This is simply for the convenience of writing (in fact, the
sum of stresses associated with different media is a non-sense).
\\
Let
\begin{equation*}
\sigma =\sum_{i=1}^{2}\sigma _{i},
\end{equation*}%
where%
\begin{equation*}
\sigma _{i\nu \gamma }=-p_{i}\delta _{\nu \gamma }-\Phi _{i\nu }\rho
_{i_{,\gamma }}-\Psi _{i\nu }s_{i_{,\gamma }} ,  \tag{18}
\end{equation*}%
with
\begin{equation*}
p_{i}=\mathcal{P}_{i}-\rho _{i}\func{div}\mathbf{\Phi }_{i}.
\end{equation*}%
The tensorial natures of entropy and density are different; that is
the reason why only $\func{div}\mathbf{\Phi }$ appears in the
pressure term. The calculus for this case appears in appendix 2. In
appendix 4, we deduce Antonov's rule related to the surface tensions
of an interface between two immiscible fluids from the expression of
the stress tensor.

For mixtures with internal capillarity ([Gatignol \& Seppecher,
1986]; [Seppecher, 1987]), only $\mathbf{\Phi }_{i}$, $i\in
\{1,2\}$, are not equal to zero. In the case of classical
compressible mixtures, terms of gradients are null and $\mathbf{\Phi
}_{i}$ and $\mathbf{\Psi }_{i}$ are zero; $\sigma _{\nu \gamma }$ is
written:
\begin{equation*}
-\mathcal{P}\delta _{\nu \gamma }
\end{equation*}%
with $\mathcal{P=}\sum_{i=1}^{2}\mathcal{P}_{i}$ defining the pressure of
the mixture.
\\
Let us remark that $\mathcal{P}_{i}$ with $i\in \{1,2\}$ does not
correspond to the classic notion of partial pressures in ideal
mixtures. This result fits with I. M\"{u}ller's paper ([1968], p.
37): according to M\"{u}ller terminology, if $\varepsilon =\rho
\alpha =\rho _{1}\alpha _{1}+\rho _{2}\alpha _{2}$ with $\alpha
_{i}=\alpha _{i}(\rho _{i},s_{i})$, such an internal energy
corresponds to a simple mixture of fluids. If $\Pi _{i}=\rho
_{i}^{2}\alpha _{i_{,\rho _{i}}}(\rho _{i},s_{i})$, then
\begin{equation*}
\mathcal{P}_{1}=\Pi _{1}+\frac{\rho _{1}\rho _{2}}{\rho }(\alpha _{1}-\alpha
_{2})
\end{equation*}%
and
\begin{equation*}
\mathcal{P}_{2}=\Pi _{2}+\frac{\rho _{1}\rho _{2}}{\rho }(\alpha _{2}-\alpha
_{1});
\end{equation*}%
with assuming that for every point in the mixture all the
constituents have the same temperature $\theta _{i}=\theta _{0}$,
then, $\mathcal{P=}$ $\Pi _{1}+\Pi _{2}$ and $\Pi _{i}$ represents
the partial pressure of constituent $i$ [Bruhat 1968].

\textit{\quad Example of binary system with complete miscibility:
liquid-vapor equilibrium}

At a given temperature, the numerous constitutive equations represent $%
\mathcal{P}$ as a function of densities of the constituents ([Soave,
1972]; [Simonet \& Behar, 1976]; [Peng \& Robinson, 1976]). These
equations, very useful in petroleum industry, derive from subtle
calculus on mixtures of Van der Waals fluids. Taking into account
Eq. (15), they allow the study of a liquid bulk and a vapour bulk
for two fluids that can be completely mixed.
\\
A plane interface lies between two phases of a two-fluid mixture in
isothermal equilibrium. This is an unidimensional problem with reference to
the direction perpendicular to the interface. Body forces being neglected,
Eq. (17) can be written:%
\begin{equation*}
\frac{d\sigma }{dz}=0.
\end{equation*}%
Vectors $\mathbf{\Phi }_{i}$ and $\mathbf{\Psi }_{i}$ are zero in
the bulks in which $\mathcal{P}_{l}=\mathcal{P}_{v}=\mathcal{P}_{0}$
($l$ and $v$ refer to the bulks with the same pressure $P_0$).
\\
For equilibrium, eight variables $\rho _{il}$, $\rho _{iv}$, $s_{il}$, $%
s_{iv}$ with $i\in \{1,2\}$ determine the properties of the two bulks. They
are subject to six relations:%
\begin{equation*}
\begin{array}{c}
g_{1l}=g_{1v}, \\
g_{2l}=g_{2v}, \\
\mathcal{\theta }_{1l}=\mathcal{\theta }_{1v}=\mathcal{\theta }_{2l}=%
\mathcal{\theta }_{2v},%
\end{array}%
\end{equation*}%
and%
\begin{equation*}
\mathcal{P}_{l}=\mathcal{P}_{v}.
\end{equation*}%
This fits with Gibbs law of phases that gives the system as "divariant"
[Rocard, 1967]. Two additional conditions, $\mathcal{\theta }_{il}=\mathcal{%
\theta }_{iv}=\mathcal{\theta }_{0}$ $(i\in \{1,2\})$ and $\mathcal{P}_{l}=%
\mathcal{P}_{v}=\mathcal{P}_{0},$ set an equilibrium of the mixture.

\subsection{Equation of motion and equation of energy for a thermocapillary
mixture}

\textit{\quad  Conservative motions}

Some properties of mixtures are directly deduced from the equation
(17) of motion.

Let us define%
\begin{equation*}
\begin{array}{l}
\mathbf{M}_{i}=\mathbf{\Gamma }_{i}-\theta _{i}\func{grad}s_{i}+\func{grad}%
(h_{i}+\Omega _{i}), \\
B_{i}=\displaystyle\frac{d\rho _{i}}{dt}+\rho _{i}\func{div}\mathbf{V}_{i}, \\
S_{i}=\displaystyle\rho _{i}\theta _{i}\frac{ds_{i}}{dt}, \\
E_{i}=\displaystyle\frac{\partial e_{i}}{\partial t}+\func{div}\left[(e_{i}-\sigma _{i})%
\mathbf{V}_{i}\right]-\func{div}\mathbf{U}_{i}-\rho _{i}\frac{\partial \Omega _{i}%
}{\partial t},%
\end{array}%
\end{equation*}%
\qquad\ \qquad\ with%
\begin{equation*}
\begin{array}{lll}
e_{i}=\displaystyle\frac{1}{2}\,\rho
_{i}\mathbf{V}_{i}^{2}+\varepsilon _{i}+\rho
_{i}\Omega _{i}, &  & \displaystyle\varepsilon _{i}=\frac{\rho _{i}\varepsilon }{\rho }%
\end{array}%
\end{equation*}%
and
\begin{equation*}
\begin{array}{c}
\mathbf{U}_{i}=\displaystyle\frac{d\rho _{i}}{dt}\,\mathbf{\Phi
}_{i}+\frac{ds_{i}}{dt}\, \mathbf{\Psi }_{i} \\  \displaystyle
\left( \frac{d\rho _{i}}{dt}=\frac{\partial \rho _{i}}{\partial t}+\frac{%
d\rho _{i}}{d\emph{\textbf{x}}}\mathbf{V}_{i}\text{ \ \ and \ \ }\frac{ds_{i}}{dt}=%
\frac{\partial s_{i}}{\partial t}+\frac{\partial s_{i}}{\partial \emph{\textbf{x}}}%
\mathbf{V}_{i}\right) .%
\end{array}%
\end{equation*}

\begin{theorem}
For a thermocapillary internal energy and for any motion of the mixture, the
relation
\begin{equation*}
\sum_{i=1}^{2}E_{i}-\rho _{i}\mathbf{M}_{i}^{\ast }\mathbf{V}i-\left(\frac{1}{2}%
\mathbf{V}_{i}^{2}+h_{i}+\Omega _{i}\right)B_{i}-S_{i}=0  \tag{19}
\end{equation*}%
is  identically satisfied.
\end{theorem}
(Proof in appendix 3). Limit cases (the isentropic and isothermal
motion of Sec. 2) yield  the two corollaries:

\begin{corollary}
- Any isentropic motion of an thermocapillary mixture satisfies the equation
of balance of energy:%
\begin{equation*}
\sum_{i=1}^{2}\frac{\partial e_{i}}{\partial
t}+\func{div}[(e_{i}-\sigma
_{i})\mathbf{V}_{i}]-\func{div}\mathbf{U}_{i}-\rho
_{i}\frac{\partial \Omega _{i}}{\partial t}=0 .  \tag{20}
\end{equation*}
\end{corollary}
This results from the simultaneity of equations $S_{i}=0$, $B_{i}=0$ and $%
\mathbf{M}_{i}=0$.

\begin{corollary}
- For any conservative motion of a thermocapillary perfectly heat conducting
mixture, Eq. (13) is equivalent to:%
\begin{equation*}
\sum_{i=1}^{2}\frac{\partial f_{i}}{\partial
t}+\func{div}[(f_{i}-\sigma
_{i})\mathbf{V}_{i}]-\func{div}\mathbf{U}_{i}-\rho
_{i}\frac{\partial \Omega _{i}}{\partial t}=0   \tag{21}
\end{equation*}%
with $f_{i}=\displaystyle\frac{1}{2}\,\rho
_{i}\mathbf{V}_{i}^{2}+\varphi _{i}+\rho _{i}\Omega _{i}$ and
$\varphi _{i}=\varepsilon _{i}-\rho _{i}\theta _{i}s_{i} $
corresponding to the \textrm{specific free energy} of constituent
$i$ at temperature $\theta _{0}$.
\end{corollary}

Equation (21) is called the \textit{equation of balance of free
energy}.

\textit{\quad Dissipative motions}

If we try to introduce dissipative phenomena, while keeping the form
of the equations, we may only add an irreversible stress tensor
$\sigma _{Ii}$ for each constituent. So, for each fluid, $(i\in
\{1,2\})$, supplementary energy flux vectors $\func{div}(\sigma
_{Ii}\,\mathbf{V}_{i})$ and dissipative
forces $tr(\sigma _{Ii}\,\mathbf{\Delta }_{i})$ have additive properties ($%
\Delta _{i}$ is the deformation tensor of constituent $i$) and are
introduced and summed. The only assumption for functions of
dissipation, $\sum_{i=1}^{2}tr(\sigma _{Ii}\,\mathbf{\Delta
}_{i})\geqq 0$, leads to a Clausius-Duhem inequality. So, the model
fits with the second law of thermodynamics independently of any
constitutive behavior for $\sigma _{I1}$ and $\sigma _{I2}$ (Chapman
and Cowling [1970]).

Equation of motion (17) can be written:
\begin{equation*}
\rho \,\mathbf{\Gamma }=\func{div}(\sigma +\sigma _{I})-\rho\,
\func{grad}\Omega
\end{equation*}%
(as in Eq.(17), we are led to formally writing $\sigma _{I}=\sigma
_{I1}+\sigma _{I2}$).

Afterwards, the heat flux vector will be denoted by $q$ and heat supply by $%
r $. It follows,

\begin{corollary}
- For any motion of a thermocapillary mixture, the \emph{equation of energy}
\begin{equation*}
\sum_{i=1}^{2}\frac{\partial e_{i}}{\partial t}+\func{div}[(e_{i}-\sigma
_{i}-\sigma _{Ii})\mathbf{V}_{i}]-\func{div}\mathbf{U}_{i}-\rho _{i}\frac{%
\partial \Omega _{i}}{\partial t}+\func{div}q-r=0   \tag{22}
\end{equation*}%
is equivalent to the \emph{equation of entropy}:%
\begin{equation*}
\sum_{i=1}^{2}\rho _{i}\theta _{i}\frac{ds_{i}}{dt}-tr(\sigma _{Ii}\mathbf{%
\Delta }_{i})+\func{div}q-r=0.
\end{equation*}
\end{corollary}

Assuming that the work of dissipative forces satisfies
$\sum_{i=1}^{2}tr(\sigma
_{Ii}\mathbf{\Delta }_{i})\geqq 0$, we obtain a generalized form of \textit{%
Planck's} \textit{inequality}:%
\begin{equation*}
\sum_{i=1}^{2}\rho _{i}\theta _{i}\frac{ds_{i}}{dt}+\func{div}q-r\geqq 0.
\end{equation*}%
At any point of the fluid, each constituent is supposed to have the
same temperature (not necessarily uniform). This requires that the
time for relaxation of molecules is short in comparison with the
time characterizing the flow. Then,
\begin{equation*}
\theta =\theta _{i}.
\end{equation*}%
By use of \textit{Fourier's law} in the form:%
\begin{equation*}
q.\func{grad}\theta \leqq 0,
\end{equation*}%
we deduce a generalization of \textit{Clausius-Duhem's inequality}:%
\begin{equation*}
\sum_{i=1}^{2}\rho _{i}\frac{ds_{i}}{dt}+\func{div}\left( \frac{q}{\theta }%
\right) -\frac{r}{\theta }\geqq 0.
\end{equation*}%
This Clausius-Duhem's inequality is a form of the second law of
thermodynamics. Let us note that some authors separate $q$ and $r$ in $q_{i}$ and $%
r_{i}$; this does not seem necessary.

\bigskip

\begin{center}
\textbf{Appendix 1}
\end{center}

\bigskip

\textbf{Equations of motion for mixtures of perfect fluids}

\textit{\quad Isentropic motion}

Let us denote by $\mathbf{\xi}_i$ the variation $\delta
_{i}\emph{\textbf{X}}_{i}$, we
deduce from the two relations in lagrangian variables [Gouin, 1987]:%
\begin{equation*}
\begin{array}{lll}\displaystyle
\delta _{i}\mathbf{V}_{i}=-\mathbf{F}_{i}\frac{\partial \mathbf{\xi}_i}{\partial t}%
, &  &\displaystyle \delta _{i}\rho _{i}=\rho _{i}\,{\func{div}}_{oi}\,\mathbf{\xi}_i+\frac{\rho _{i}%
}{\rho _{oi}}\frac{\partial \rho _{oi}}{\partial
\emph{\textbf{X}}_{i}}\,\mathbf{\xi}_i  \tag{23}
\end{array}%
\end{equation*}%
and%
\begin{equation*}
\delta _{i}s_{i}=\frac{\partial s_{i}}{\partial \emph{\textbf{X}}_{i}}\,\delta _{i}%
\emph{\textbf{X}}_{i},
\end{equation*}%
where $\displaystyle\frac{\partial }{\partial
\emph{\textbf{X}}_{i}}$ is the linear form associated with the
gradient and $\func{div}_{oi}$ is the divergence operator on
$\mathcal{D}_{oi}$.
\\
Variations of entropies satisfy relation (9). With assuming that
terms on the
edge of $\mathcal{W}$ are zero, we obtain:%
\begin{eqnarray*}
\delta _{i}a &=&\int_{t_{1}}^{t_{2}}\int_{\mathcal{D}_{t}}\left\{ \left(\frac{1}{2%
}\mathbf{V}_{i}^{\ast }\mathbf{V}_{i}-\varepsilon _{,\rho _{i}}-\Omega
_{i}\right)\delta _{i}\rho _{i}+\rho _{i}\mathbf{V}_{i}^{\ast }\delta _{i}\mathbf{V%
}_{i}-\varepsilon _{,s_{i}}\delta _{i}s_{i}\right. \\
&&-\left[ \sum_{p=1}^{n-1}(-1)^{p}(\varepsilon _{,\rho _{i,_{\gamma
_{1},...,\gamma _{p})}}})_{,\gamma _{1},...,\gamma _{p}}\right] \delta
_{i}\rho _{i} \\
&&\left. -\left[ \sum_{p=1}^{n-1}(-1)^{p}(\varepsilon
_{,s_{i,_{\gamma _{1},...,\gamma _{p})}}})_{,\gamma _{1},...,\gamma
_{p}}\right] \delta _{i}s_{i}\right\} dv dt, \quad \rm{or}
\end{eqnarray*}%
\begin{equation*}
\delta _{i}a=\int_{t_{1}}^{t_{2}}\int_{\mathcal{D}_{oi}}\rho
_{oi}\left[
\mathbf{R}_{i}\,{\func{div}}_{oi}\,\mathbf{\xi}_i+\frac{\mathbf{R}_{i}}{\rho _{oi}}\frac{%
\partial \rho _{oi}}{\partial \emph{\textbf{X}}_{i}}\,\mathbf{\xi}_i-\mathbf{V}_{i}^{\ast }%
\mathbf{F}_{i}\frac{\partial \mathbf{\xi}_i}{\partial t}-\theta
_{i}\frac{\partial s_{i}}{\partial
\emph{\textbf{X}}_{i}}\,\mathbf{\xi}_i\right] dv_{oi}dt
\end{equation*}%
with $\mathbf{R}_{i}$ denotes $\displaystyle\frac{1}{2}\mathbf{V}_{i}^{\ast }\mathbf{V}%
_{i}-h_{i}-\Omega _{i}$.
\\
But,%
\begin{equation*}
{\func{div}}_{oi}(\rho _{oi}\mathbf{R}_{i}\mathbf{\xi}_i)=\rho _{oi}\mathbf{R}_{i}%
{\func{div}}_{oi}\,\mathbf{\xi}_i+\mathbf{R}_{i}\frac{\partial \rho
_{oi}}{\partial \emph{\textbf{X}}_{i}}\,\mathbf{\xi}_i+\rho
_{oi}\frac{\partial \mathbf{R}_{i}}{\partial
\emph{\textbf{X}}_{i}}\,\mathbf{\xi}_i.
\end{equation*}%
So,%
\begin{equation*}
\delta _{i}a=\int_{t_{1}}^{t_{2}}\int_{\mathcal{D}_{oi}}\rho _{oi}\,\left[ -\frac{%
\partial \mathbf{R}_{i}}{\partial \emph{\textbf{X}}_{i}}+\frac{\partial }{\partial t%
}(\mathbf{V}_{i}^{\ast }\mathbf{F}_{i})-\theta _{i}\frac{\partial s_{i}}{%
\partial \emph{\textbf{X}}_{i}}\right] \mathbf{\xi}_i\,dv_{oi}dt,
\end{equation*}%
(terms given by integration on the edge of $\mathcal{D}_{oi}$ are
zero).

Finally, for each constituent,%
\begin{equation*}
\frac{\partial }{\partial t}(\mathbf{V}_{i}^{\ast }\mathbf{F}_{i})=\frac{%
\partial \mathbf{R}_{i}}{\partial \emph{\textbf{X}}_{i}}+\theta _{i}\frac{\partial
s_{i}}{\partial \emph{\textbf{X}}_{i}},
\end{equation*}%
and by using of the relation%
\begin{equation*}
\frac{\partial }{\partial t}(\mathbf{V}_{i}^{\ast
}\mathbf{F}_{i})=\mathbf{\Gamma}
_{i}^{\ast }\mathbf{F}_{i}+\mathbf{V}_{i}^{\ast }\frac{\partial \mathbf{V}%
_{i}}{\partial \emph{\textbf{X}}_{i}},
\end{equation*}%
we obtain
\begin{equation*}
\mathbf{\Gamma }_{i}^{\ast }\mathbf{F}_{i}+\frac{\partial }{\partial
\emph{\textbf{X}}_{i}}(\frac{1}{2}\mathbf{V}_{i}^{\ast
}\mathbf{V}_{i})=\frac{\partial
\mathbf{R}_{i}}{\partial \emph{\textbf{X}}_{i}}+\theta _{i}\frac{\partial s_{i}}{%
\partial \emph{\textbf{X}}_{i}}
\qquad {\rm or}\qquad
\mathbf{\Gamma }_{i}^{\ast }=\theta _{i}\frac{\partial s_{i}}{\partial \emph{\textbf{x}}}-%
\frac{\partial }{\partial \emph{\textbf{x}}}(h_{i}+\Omega _{i}).
\end{equation*}%
This leads to the vectorial form:%
\begin{equation*}
\mathbf{\Gamma }_{i}=\theta
_{i}\func{grad}s_{i}-\func{grad}(hi+\Omega _{i})  \tag{11}
\end{equation*}

\begin{center}
\bigskip

\textbf{Appendix 2}
\end{center}

\textbf{Equation of motion of thermocapillary mixtures}

\bigskip

Equation (11) implies%
\begin{equation*}
\sum_{i=1}^{2}\rho _{i}\Gamma _{i\gamma }+\rho _{i}\Omega _{i,_{\gamma
}}=\sum_{i=1}^{2}\varepsilon _{,s_{i}}s_{i,_{\gamma }}-(\varepsilon
_{,s_{i,_{\nu }}})_{,_{\nu }}s_{i,_{\gamma }}-\rho _{i}(\varepsilon _{,\rho
_{i}})_{,_{\gamma }}+\rho _{i}[(\varepsilon _{,\rho _{i,_{\nu }}})_{,_{\nu
}}]_{,_{\gamma }}.
\end{equation*}%
By noting that%
\begin{equation*}
\varepsilon _{,\gamma }=\sum_{i=1}^{2}\varepsilon _{,s_{i}}s_{i,_{\gamma
}}+(\varepsilon _{,s_{i,\nu }})_{,_{\nu }}s_{i,_{\nu \gamma }}+\varepsilon
_{,\rho _{i}}\rho _{i,_{\gamma }}+\varepsilon _{,\rho _{i,\nu }}\rho
_{i,_{\nu \gamma }},
\end{equation*}%
we obtain%
\begin{eqnarray*}
&&\sum_{i=1}^{2}\rho _{i}\Gamma _{i\gamma }+\rho _{i}\Omega
_{i,_{\gamma }}  =
\\
&&\varepsilon _{,_{\gamma }}-\left\{ \sum_{i=1}^{2}\varepsilon
_{,s_{i,_{\nu }}}s_{i,_{\nu \gamma }}+\varepsilon _{,\rho _{i}}\rho
_{i,_{\gamma }}+\varepsilon _{,\rho _{i,\nu }}\rho _{i,_{\nu \gamma
}}+(\varepsilon _{,s_{i,_{\nu }}})_{,_{\nu }}s_{i,_{\gamma
}}+\right.
\\
&&\qquad\qquad\left.\rho _{i}(\varepsilon _{,\rho _{i}})_{,_{\gamma
}}-\rho _{i}[(\varepsilon _{,\rho _{i,_{\nu }}})_{,_{\nu
}}]_{,_{\gamma }}\right\} ,
\end{eqnarray*}%
or%
\begin{equation*}
\sum_{i=1}^{2}\rho _{i}\Gamma _{i\gamma }+\rho _{i}\Omega
_{i,_{\gamma }}=\sum_{i=1}^{2}\left[ \varepsilon -\rho
_{i}\varepsilon _{,\rho _{i}}+\rho _{i}(\varepsilon _{,\rho
_{i,_{\nu }}})_{,_{\nu }}\right]_{,_{\gamma }}-\left[\Phi
_{i\nu }\rho _{i_{,_{\gamma }}}+\Psi _{i\nu }s_{i_{,_{\gamma }}}\right] _{,_{\nu }}%
.
\end{equation*}%
In the case $\Omega _{1}=\Omega _{2}=\Omega $, this agrees with Eq.
(17) when $\rho\, \alpha =\varepsilon $.

\bigskip

\begin{center}
\textbf{Appendix 3}
\end{center}

\bigskip

\textbf{Proof of relation (19)}

Relation $(10^{1})$ yields%
\begin{equation*}
E_{i}=\frac{\partial e_{i}}{\partial t}+\func{div}\left[\rho _{i}\left(\frac{1}{2}%
\mathbf{V}_{i}^{2}+h_{i}+\Omega _{i}\right)\mathbf{V}_{i}\right]-\func{div}\left[\frac{%
\partial \rho _{i}}{\partial t}\mathbf{\Phi }_{i}+\frac{\partial s_{i}}{%
\partial t}\mathbf{\Psi }_{i}\right]-\rho _{i}\frac{\partial \Omega _{i}}{\partial
t}.
\end{equation*}%
On the left hand side of Eq.(19), the extraneous force potential and inertia
terms cancel out.
\\
With $e_{i},h_{i},\mathbf{\Phi }_{i}$ and $\mathbf{\Psi }_{i}$ having been
replaced by their respective values, it remains to be proved that the terms
from the internal energy also cancel out.
\\
By using the convention that terms subscripted by $i$ are \textit{summed on} $%
\{1,2\}$, the internal energy involves:

a) in $E_{i}$: $\displaystyle\frac{\partial \varepsilon }{\partial
t}+\func{div}(\rho
_{i}h_{i}\mathbf{V}_{i})-\func{div}\left(\frac{\partial \rho _{i}}{\partial t}%
\mathbf{\Phi }_{i}+\frac{\partial s_{i}}{\partial t}\mathbf{\Psi
}_{i}\right),$

b) in $\rho _{i}\mathbf{M}_{i}^{\ast }\mathbf{V}_{i}$ :\ $-\rho
_{i}\theta _{i}\func{grad}^{\ast }s_{i}\ \mathbf{V}_{i}+\rho
_{i}\func{grad}^{\ast }h_{i}\ \mathbf{V}_{i}\,,$

c) in $\displaystyle\left(\frac{1}{2}\mathbf{V}_{i}^{2}+h_{i}+\Omega
_{i}\right)B_{i}$: $\displaystyle h_{i}\left( \frac{d\rho
_{i}}{dt}+\rho _{i}\func{div}\mathbf{V}_{i}\right) ,$

d) in $S$ : $\displaystyle \rho _{i}\theta
_{i}\left(\frac{ds_{i}}{dt}\right).$

This leaves the following in the first term of (19):%
\begin{equation*}
\frac{\partial \varepsilon }{\partial t}+\func{div}(\rho _{i}h_{i}\mathbf{V}%
_{i})-\func{div}\left(\frac{\partial \rho _{i}}{\partial t}\mathbf{\Phi }_{i}+%
\frac{\partial s_{i}}{\partial t}\mathbf{\Psi }_{i}\right)
\end{equation*}
\begin{equation*}
-\rho _{i}\theta _{i}\left(%
\frac{\partial s_{i}}{\partial t}\right)-\rho _{i}\,{\func{grad}}^{\ast }h_{i}\,\mathbf{%
V}_{i}-h_{i}\left( \frac{d\rho _{i}}{dt}+\rho _{i}\func{div}\mathbf{V}%
_{i}\right) ,
\end{equation*}
or:%
\begin{equation*}
\frac{\partial \varepsilon }{\partial t}-(h_{i}+\func{div}\mathbf{\Phi }_{i})%
\frac{\partial \rho _{i}}{\partial t}-(\rho _{i}\theta _{i}+\func{div}%
\mathbf{\Psi }_{i})\frac{\partial s_{i}}{\partial t}-\mathbf{\Phi }%
_{i}^{\ast }\frac{\partial \func{grad}\rho _{i}}{\partial t}-\mathbf{\Psi }%
_{i}^{\ast }\frac{\partial \func{grad}s_{i}}{\partial t}.
\end{equation*}%
Definitions (16) of $\varepsilon $ and (10) of $h_{i}$ and $\theta _{i}$
show immediately that this expression is identically zero.

\bigskip

\begin{center}
\textbf{Appendix 4}
\end{center}

\bigskip

\textbf{Interpretation of capillary tensions associated with two immiscible
fluids}

\bigskip

We consider an interface between two immiscible fluids. The assumption that
two fluids remain absolutely separated is not reflected in physical
experiments. The two fluids interpenetrate each other in a thin layer
(thickness about few molecular diameters). Energetic reasons justify the
assumption [Brin, 1956]. The density of each constituent regularly changes
between the value in bulk $1$ and the one in bulk $2$, with a profile in
accordance with the one of a liquid-vapour interface; each density vanishes
asymptotically in the complementary phase (see \textit{Fig.2}).
\\
In the interfacial layer, density gradients are important. In the isothermal
case, the zone of mixture of two isotropic fluids can be represented with an
internal energy of the form:%
\begin{equation*}
\varepsilon =\rho \alpha (\rho _{1},\rho _{2},\beta _{1},\beta _{2},\gamma )
\end{equation*}%
with%
\begin{equation*}
\begin{array}{lllll}
\beta _{i}=(\func{grad}\rho _{i})^{2} &  & \text{and} &  & \gamma =\func{grad%
}\rho _{1}\func{grad}\rho _{2}.%
\end{array}%
\end{equation*}%
The stress tensor reduced from Eq. (18) is:%
\begin{equation*}
\sigma =\sum_{i\neq j}\left[-\rho \rho _{i}\alpha _{,_{\rho _{i}}}+\rho _{i}\func{%
div}(C_{i}\func{grad}\rho _{i}+D\func{grad}\rho _{i})\right]Id
\end{equation*}
\begin{equation*}
-C_{i}\func{grad}%
\rho _{i}\,{\func{grad}}^{\ast }\rho _{i}-D\func{grad}\rho
_{i}\,{\func{grad}}^{\ast }\rho _{j},
\end{equation*}

with $C_{i}=2\rho \alpha
_{,_{\beta _{i}}}$ and $D=\rho \alpha _{,_{\gamma }}.$
Let us denote%
\begin{equation*}
\Pi =\sum_{i\neq j}\rho \rho _{i}\alpha _{,_{\rho _{i}}}-\rho _{i}\func{div}%
(C_{i}\func{grad}\rho _{i}+D\func{grad}\rho
_{i})+C_{i}(\func{grad}\rho _{i})^{2}+D{\func{grad}}^{\ast }\rho
_{i}\,{\func{grad}}\rho _{j}.
\end{equation*}%
In each bulk, gradient terms are zero and $\Pi $ takes the values:%
\begin{equation*}
\begin{array}{lllll}
\Pi _{1}=\rho _{1}^{2}\alpha _{,_{\rho _{1}}}(\rho _{1},0,0,0,0) &  & \text{%
and} &  & \Pi _{2}=\rho _{2}^{2}\alpha _{,_{\rho _{2}}}(0,\rho _{2},0,0,0)%
\end{array}%
\end{equation*}
that correspond to the pressure in each phase of fluid.

\begin{figure}[h]
\begin{center}
\includegraphics[width=15cm]{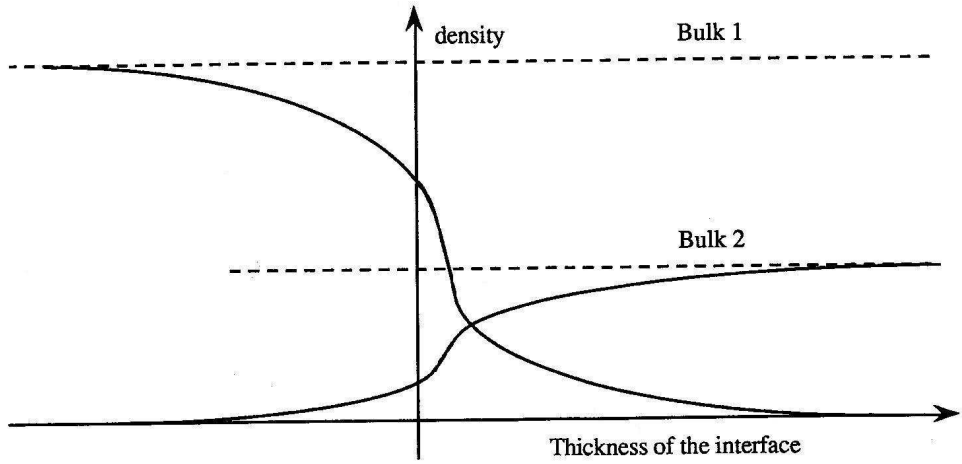}
\end{center}
\caption{Profiles of densities through the interface}\label{fig1}
\end{figure}

When the extraneous force potential are neglected,%
\begin{equation*}
\func{div}\sigma =0,
\end{equation*}%
which is the equation of the equilibrium state of the mixture.

For a flat interface normal to $\func{grad}\rho _{1}$ and $\func{grad}\rho
_{2}$, the eigenvalues of stress tensor $\sigma $ are%
\begin{equation*}
\lambda _{1}=-\Pi +C_{1}(\func{grad}\rho
_{1})^{2}+C_{2}(\func{grad}\rho _{2})^{2}+2D {\func{grad}}^{\ast
}\rho _{1}\func{grad}\rho _{2},
\end{equation*}%
(associated with the plane of interface), and%
\begin{equation*}
\lambda _{2}=-\Pi
\end{equation*}%
(associated with the direction normal to the plane of interface).
\\
In a system of coordinates suitable for the interface (\textit{i.e.}
an orthonormal system whose third direction is $z$, the stress
tensor is
written:%
\begin{equation*}
\sigma =%
\begin{bmatrix}
\lambda _{1} & 0 & 0 \\
0 & \lambda _{1} & 0 \\
0 & 0 & \lambda _{2}%
\end{bmatrix}%
.
\end{equation*}%
The equation of balance momentum in the planar interface implies:%
\begin{equation*}
\lambda _{2}=-\Pi _{0},
\end{equation*}%
where $\Pi _{0}$ denotes the pressure that must be common to the two bulks.

The line force per unit of length on the edge of the interface is (\textit{%
see Fig.} 3):%
\begin{equation*}
\mathcal{F}=\int_{z_{1}}^{z_{2}}\lambda _{1}dz=
\end{equation*}
\begin{equation*}
 -\Pi _{0}(z_{2}-z_{1})+\int_{z_{1}}^{z_{2}}\{C_{1}(\func{grad}\rho
_{1})^{2}+C_{2}(\func{grad}\rho _{2})^{2}+2D{\func{grad}}^{\ast }\rho _{1}%
\func{grad}\rho _{2}\}dz,
\end{equation*}

where $z_{2}-z_{1}$ corresponds to the physical size of the
interface.
\\
Let us note that, due to the thickness of the interface, $-\Pi
_{0}(z_{2}-z_{1})$ is negligible.
\\
Conditions on the edge are not the ones of first gradient fluid.
Nevertheless, the problem concerns a plane interface and the mean
radius of curvature $R_{m}$ is such that $R_{m}=\infty $. Therefore,
the constraint condition on the edge is in the form $n^{\ast }\sigma
$ where $n^{\ast }$ is normal to the edge of the interface [Casal,
1972].
\begin{figure}[h]
\begin{center}
\includegraphics[width=10cm]{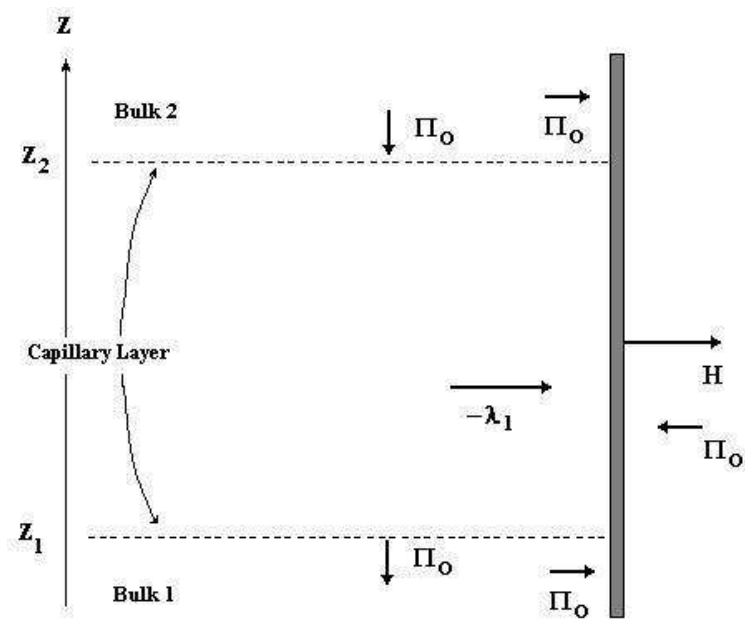}
\end{center}
\caption{Interpretation of the mixture surface tension}\label{fig1}
\end{figure}
Let us note:%
\begin{equation*}
\begin{array}{lllll}
H_{i}=\displaystyle\int_{z_{1}}^{z_{2}}\{C_{i}(\func{grad}\rho _{i})^{2}\}\,dz &  & \text{%
et} &  & \displaystyle H_{1.2}=\int_{z_{1}}^{z_{2}}2D{\func{grad}}^{\ast }\rho _{1}\func{%
grad}\rho _{2}\,dz.%
\end{array}%
\end{equation*}%
(The integrals converge in the two bulks).
The line force per unit of length on the edge is:%
\begin{equation*}
H=H_{1}+H_{2}+H_{1.2}.
\end{equation*}
Here, $H$ represents the surface tension of the flat interface in
its
equilibrium state expressed as a function of the surface tensions $H_{1}$ and $%
H_{2}$ of each bulk and a supplementary term $H_{1.2}$ associated
with the effect of the mixture. This is the expression of the
surface tension of an interface between two immiscible fluids known
as Antonov's rule [Emschwiller, 1964].

\medskip

\begin{center}
{\large References} \end{center} \footnotesize

Atkin R.J., Craine R.E., 1976, Continuum theories of mixtures: basic
theory and historical development, \textit{Q. J. Mech. Appl. Math.}, \textbf{%
29}, t. 2, 209-243.

Barrere M., Prud'homme R., 1973, \textit{Equations fondamentales de l'a%
\'{e}rothermochimie}, Masson, Paris.

Bedford A., Drumheller D.S., 1978, A variational theory of
immiscible mixtures, \textit{Arch. Rat. Mech. Anal.}, \textbf{68},
37-51.

Bedford A., Drumheller D.S., 1983, Recent advances. Theories of
immiscible and structured mixtures. \textit{Int. J. Engng. Sci.},
\textbf{21}, 8, 863-960.

Berdichevskii V.L., 1966, Construction of models of continuous media
by means of the variational principle, \textit{J. Appl. Math. Mech.}
(translation of Soviet journal P.M.M.), \textbf{30}, 510-530.

Bowen R.M., 1979, A theory of constrained mixtures with multiple
temperatures, \textit{Arch. Rat. Mech. Anal.}, \textbf{70}, 235-250.

Brin A., 1956, Contribution \`{a} l'\'{e}tude de la couche
capillaire et de la pression osmotique, \textit{Thesis}, Paris.

Bruhat G., 1968, \textit{Thermodynamique}, 6th. Ed., Masson, Paris,
188-192.

Cahn J.W., Hilliard J.E., 1959, Free energy of a non-uniform system,
\textit{J. Chem. Phys.}, \textbf{31}, 688-699.

Casal P., 1966,  Principes variationnels en fluide compressible et en magn%
\'{e}todynamique des fluides, \textit{J. M\'ecanique,} \textbf{5},
150-161.

Casal P., 1972, Equations \`{a} potentiels en magn\'{e}todynamique
des fluides, \textit{C.R. Acad. Sci. Paris}, \textbf{274}, A,
806-808.

Casal P., Gouin H., 1985 \textit{a}, Connection between the energy
equation and the motion equation in Korteweg's theory of
capillarity, \textit{C.R. Acad. Sci. Paris}, \textbf{300}, II,
231-234.

Casal P., Gouin H., 1985 \textit{b}, Kelvin's theorems and potential
equations in Korteweg's theory of capillarity, \textit{C.R. Acad. Sci. Paris}%
, \textbf{300}, II, 301-304.

Casal P., Gouin H., 1988 \textit{a}, Equations of motion of
thermocapillary fluids, \textit{C.R. Acad. Sci. Paris},
\textbf{306}, II, 99-104.

Casal P., Gouin H., 1988 \textit{b}, A representation of
liquid-vapour interfaces by using fluids of second grade,
\textit{Ann. Phys.}, colloque 2, Suppl. 3, \textbf{13}, 3-12.

Casal P., Gouin H.,  in: arXiv:0803.3160  \& 1989, Invariance
properties of inviscid fluids of
grade \textit{n}, in PDEs and continuum models of phase transitions, \textit{%
Lect. Notes Phys.}, \textbf{344}, 85-98, Springer

Chapman S., Cowling T.G., 1970, \textit{The Mathematical Theory of
Non-Uniform Gases}, Cambridge University Press.

Ding-Yu Peng, Robinson D.B., 1976, A new two-constant equation of
state, Ind. , \textit{Engng. Chem. Fundam.}, \textbf{15}, 1, 59-64.

Drew D.A., 1983, Mathematical modelling of two-phase flow,
\textit{Ann. Rev. Fluid Mech.}, \textbf{15}, 261-291.

Drumheller D.S., Bedford A., 1979, On the mechanics and
thermodynamics of fluid mixtures, \textit{Arch. Rat. Mech.}
\textit{Anal.}, \textbf{71}, 345-355.

Drumheller D.S., Bedford A., 1980, A thermomechanical theory for
reacting immiscible mixtures, \textit{Arch. Rat.} \textit{Mech.
Anal.}, \textbf{73}, 257-284.

Dunn J.E., Serrin J., 1985, On the thermodynamics of interstitial
working, \textit{Arch. Rat. Mech. Anal.}, \textbf{88}, 95-133.

Eglit M.E., 1965, A generalization of the model of an ideal
compressible fluid, \textit{J. Appl. Math Mech.} (translation of
Soviet journal P.M.M.), \textbf{29}, 2, 351-354.

Emschwiller G., 1964, \textit{Chimie Physique}, P.U.F., Paris.

Gatignol R., Seppecher P., 1986, Modelisation of fluid-fluid
interfaces with material properties, \textit{J. M\'ec. Th\'{e}or}.
\textit{Appl.}, special issue, 225-247.

Gouin H., 1976, Noether theorem in fluid mechanics, \textit{Mech. Res. Comm.}%
, \textbf{3}, 151-155.

Gouin H., 1978, Etude g\'{e}om\'{e}trique et variationnelle des
milieux continus, \textit{thesis}, University of Aix- Marseille 1.

Gouin H., 1981, Lagrangian representation and invariance properties
of perfect fluid flows, \textit{Res. Notes Math.}, \textbf{46},
Pitman, London, 128-136.

Gouin H., 1981, Exemples de mouvements de fluides parfaits non
conservatifs, \textit{J. M\'{e}c.}, \textbf{20}, 273-287.

Gouin H., 1987, Thermodynamic form of the equation of motion for
perfect fluids of grade $n$, \textit{C.R. Acad. Sci.}
\textit{Paris}, \textbf{305}, II, 833-838.

Green A.E., Naghdi P.M., 1965, A dynamic theory of interacting
continua, \textit{Int. J. Engng Sci.}, \textbf{3}, 231-241.

Ishii M., 1975, \textit{Thermo-fluid dynamic theory of two-phase
flow}, Eyrolles, Paris.

M\"{u}ller I., 1967, On the entropy inequality, \textit{Arch. Rat.
Mech. Anal.}, \textbf{27}, 118-141.

M\"{u}ller I., 1968, Theory of mixtures of fluids, \textit{Arch.
Rat. Mech. Anal}., \textbf{28}, 1-38.

Nunziato J.W., Walsh E.K., 1980, On ideal multiphase mixtures with
chemical reactions and diffusion, \textit{Arch.} \textit{Rat. Mech.
Anal.}, \textbf{73}, 285-331.

Rocard Y., 1967, \textit{Thermodynamique}, Masson, Paris, 52 and
143-157.

Rowlinson J.S., Swinton F.L., 1982, \textit{Liquids and Liquid
Mixtures}, Butterworth, London.

Sampaio R., Williams W.O., 1977, On the viscosities of liquid
mixtures, \textit{J. Appl. Math. Phys.}, \textbf{28}, 607-613.

Seppecher P.,1987, Etude d'une mod\'{e}lisation des zones
capillaires fluides, \textit{Thesis}, University of Paris 6.

Serrin J., 1959, Mathematical principles of classical fluid
mechanics, \textit{Encyclopedia of Physics}, VIII/1, Springer,
Berlin, 144-150.

Simonet R., Behar E., 1976, A modified Redlich-Kwong equation of
state for accurately representing pure components data,
\textit{Chem. Engng. Sci}., \textbf{31}, 37-43.

Soave G., 1972, Equilibrium constants from a modified Redlich-Kwong
equation of state, \textit{Chem. Engng. Sci.}, \textbf{27},
1197-1203.

Truesdell C., 1957, Sulle basi della termomeccanica, \textit{Rend.
Accad. Naz. Lincei}, \textbf{8}, 158-166.

Truesdell C., 1965, The rational mechanics of material,
\textit{Continuum Mechanics II}, Gordon and Breach, New York,
293-305.

Vignes-Adler M., Brenner H., 1985, A micromechanical derivation of
the differential equations of interfacial statics, \textit{J.
Collo\"{\i}d Interface Sci.}, \textbf{103}, 1, 11-44.

Williams W.O., 1973, On the theory of mixtures, \textit{Arch. Rat.
Mech. Anal.}, \textbf{51}, 239-260.

\end{document}